%% file: b630.tex
\documentclass[twocolumn,aps,showpacs,preprintnumbers,amsmath,amssymb,superscriptaddress,]{revtex4}

\usepackage{graphicx}
\usepackage{dcolumn}
\usepackage{bm}

\begin{document}


\title{Measurement of the Ratio of \boldmath{$b$} Quark Production Cross 
Sections in \boldmath{$\overline{p}p$} Collisions \\
at \boldmath{$\sqrt{s}=630$~GeV} and \boldmath{$\sqrt{s}=1800$}~GeV}

\input{authors.tex}

\date{\today}

\begin{abstract}
We report a measurement of the ratio of the bottom quark 
production cross section
in antiproton-proton collisions at $\sqrt{s}=630$~GeV to 1800~GeV 
using bottom quarks with transverse momenta greater than 10.75~GeV
identified through their semileptonic decays
and long lifetimes.  The measured ratio
$\sigma(630)/\sigma(1800) = 0.171 \pm .024 \pm .012$ is in 
good agreement with next-to-leading order (NLO) quantum 
chromodynamics (QCD).
\end{abstract}

\maketitle

\section{Introduction}

    Hadroproduction of heavy quarks, such as the bottom (or $b$) quark, 
at proton-antiproton colliders is an area where one expects perturbative 
quantum chromodynamics (QCD) to provide accurate and reliable predictions. 
Because $b$ quarks are light enough to be produced in sufficient 
quantities to enable high statistics measurements (unlike the heavier 
top quarks at the present time), they provide an excellent arena in which 
to test these predictions. It therefore came as a surprise that 
Tevatron measurements \cite{CDFBXsec, CDFBXsec2, CDFBXsec3, CDFBXsec4,
D0BXsec1, D0BXsec2, D0BXsec3} of the $b$-quark cross section 
in antiproton-proton collisions at $\sqrt{s}=1800$~GeV 
were substantially larger (roughly a factor of two) 
than predicted by next-to-leading order (NLO) QCD , 
particularly since the UA1 measurements at $\sqrt{s}=630$~GeV did not 
seem to show such a marked departure from prediction \cite{NDE-diff}.

    This disagreement could indicate that NLO QCD is insufficient and
that higher order calculations are needed.  It could
indicate that our heavy quark fragmentation models are 
insufficient, such as suggested in the paper of Cacciari, Greco 
and Nason \cite{Cacciari} which discusses improvements in theoretical 
predictions from resummation and altering fragmentation functions. It 
could also be explained by more exotic processes.  For example, Berger 
{\it et al.} \cite{Berger} propose gluino
pair production with a subsequent decay into a bottom quark and a light
bottom squark.  Since the assumed gluino mass is larger than the mass
of the $b$ quark, this process would turn on more slowly with energy
than pure QCD production of $b\overline{b}$ pairs.  This new physics
process will depress the ratio of the $b$-quark cross section at 630~GeV
relative to 1800~GeV by of order 10\%.  

    To address this apparent discrepancy, the Tevatron ran for nine days 
at an energy of $\sqrt{s}=630$~GeV to provide a sample of $b$ quarks 
produced at this energy.  Rather than calculating the 
absolute cross section at both energies and comparing, we chose to calculate 
the ratio of cross sections at the two energies.  Both experimentally and 
theoretically, many systematic uncertainties partially or completely divide out.  
In particular, the largest theoretical uncertainty is the choice 
of scale, and in predicting the ratio a consistent scale must be 
chosen at both energies: this reduces  the theoretical uncertainty 
from a factor of two to approximately 15\% for the ratio.

    This analysis identifies $b$-quark candidates by searching for long-lived 
particles with a muon as a decay product, and from the ratio of the number
of candidate events at the two energies, we compute the ratio of 
cross sections. While a differential cross section 
with respect to transverse momentum ($d\sigma(b)/dp_T$) 
would provide the best comparison with
theory, we have neither the number of events nor the $p_T$ resolution
to make a differential measurement. Instead we 
report the $b$-quark cross section above a 
minimum transverse momentum, $p_T(\rm{min})$.  We adopt the 
convention that $p_T(\rm{min})$ will be chosen so that 90\% of our 
reconstructed and identified $b$ quarks have a larger transverse 
momentum: for this analysis that is 10.75~${\rm GeV}/c$.

   In this analysis, we make the assumption that the fragmentation,
decay, and detector response to a $b$ quark of a given $p_T$ is the
same at the two energies.  Certainly the decays should be the same.
In principle, there might be a difference in fragmentation between 
630~GeV and 1800~GeV due to the difference in velocities of the proton 
remnant.  It is common to use Peterson \cite{Peterson}
fragmentation (developed for  $e^+e^-$
collisions) in $\overline{p}p$ collisions, and one would expect that
any energy-dependent fragmentation change would be smaller than
the error introduced in going from lepton to hadron colliders.  Additionally,
any difference should be at its minimum for $b$'s at central rapidity (measured
in this analysis) because they are farthest from the forward-going proton 
and antiproton remnants.

     The CDF detector is described in detail elsewhere \cite{CDFNIM}; a brief 
discussion follows. In the CDF detector, a 51 cm long silicon vertex detector 
(SVX) \cite{SVXRef}, located immediately outside the beampipe, provides 
precise track reconstruction in the plane transverse to the beam and
is used to identify secondary vertices that can be produced 
by $b$ and $c$ quark decays. Because $p\overline{p}$ interactions are 
spread along the beamline with a standard deviation of about 30 cm, 
slightly more than half of the events originate from primary vertices inside
the SVX fiducial region (this fraction is a function of beam energy). The momentum 
of charged particles is measured in the central tracking chamber 
(CTC), which sits inside a 1.4 T superconducting solenoidal magnet.  
Outside the CTC are electromagnetic and hadronic calorimeters 
arranged in a  projective tower geometry, covering the pseudorapidity region 
$|\eta| < 4.2$ \cite{Phi}.   Surrounding the calorimeters, drift chambers in 
the region $|\eta| < 1.0$ provide muon identification.  
In this analysis, we restrict ourselves to muons in the most central 
region ($|\eta|<0.6$), requiring muons detected in both the inner 
central muon chambers (CMU), located behind approximately five
interaction lengths of material, and the outer central muon upgrade
chambers (CMP) behind an additional 60~cm of steel.

\section{Data Selection}

   Our goal was to make the two datasets (630~GeV and 1800~GeV) as similar as
possible.  All the data were collected between December 1995 and 
February 1996.  Therefore, changes to the detector configuration and
time-dependent effects were minimized.

    Both online and offline event selections were identical for the two
beam energies.  A three-level trigger selected events with a
high transverse momentum muon for this analysis.  A muon was 
identified by requiring a match between the extrapolated track 
as reconstructed in the CTC and track segments reconstructed 
in the muon chambers, taking into account multiple scattering 
of the muon.  At Level 1, events were selected online by having 
at least one identified short track  (called a ``stub", 
having at least two hits out of four possible) in the CMU muon chambers 
with confirming hits in the outer CMP muon chambers. At Level 2,
events were required to have a 4.7~${\rm GeV}/c$ $p_T$ two-dimensional 
($r$-$\phi$) track in the central tracker pointing at a stub in the CMU. 
At Level 3, events were selected that had a good 3-dimensional track 
with $p_T > 4.5$~${\rm GeV}/c$ pointing at muon stubs with at least three hits
in both the CMU and CMP chambers.  Offline, the muon candidate 
was required to pass tight track-stub matching requirements:
the momentum-dependent matching $\chi^2$ must have been less 
than 9 in the $x$-direction for both  CMU and CMP, and must have been 
less than 12 in the $z$-direction for CMU.  The $\chi^2$ variables were
calculated for one degree of freedom. The muon track
was required to have $p_T > 5.0$~${\rm GeV}/c$ as well as to have at least 3 
(of 4 possible) hits in the SVX.  For muons with $p_T$ 
above 6~${\rm GeV}/c$, 
the trigger efficiency is essentially constant (variation less 
than 1\% with $p_T$). Monte Carlo calculations indicate 
90 percent of the $b$ quarks passing these requirements have 
transverse momenta above 10.75~${\rm GeV}/c$.

   Much of the 1800~GeV sample had the 4.7~${\rm GeV}/c$ Level 2 muon trigger 
dynamically prescaled.  At high luminosities, these triggers were 
run with a very high prescale factor (100 or more), and as the luminosity 
decreased, the prescale factors were lowered until the trigger 
ran with no prescale.  This strategy maximizes the number of events 
recorded to tape, but complicates the calculation of the 
live luminosity.  We elected to use the data itself to make this
calculation.  (Run-by-run bookkeeping yields a consistent result.)  
We looked at muon events that passed an unprescaled 12~${\rm GeV}/c$
muon trigger and we subjected them to the same offline cuts
used for our sample, except that the minimum $p_T$ was required to be
15~${\rm GeV}/c$. Every one of these events 
should have passed the unprescaled 4.7~${\rm GeV}/c$ muon 
trigger, so this sample
allows us to determine the effective prescale factor.
We have 3943 such events, of which 1282 pass the prescaled trigger: 
our raw luminosity must therefore be multiplied by a prescale correction of
$0.3250 \pm 0.0075$.  Applying this effective prescale 
and the luminosity systematic uncertainty of 4.2\% \cite{luminosity}, 
we get an effective integrated luminosity of 
$623 \pm 30$~${\rm nb^{-1}}$ at 1800~GeV. 

  At 630~GeV, this trigger had no prescale applied, and we collected an 
integrated luminosity  of $582 \pm 24$~${\rm nb^{-1}}$.
  
   The uncertainties on the integrated luminosities are independent, so the 
uncertainty on the ratio is straightforward to calculate.  We 
obtain the ratio:

\[  \frac
{{\cal L}(630)}
{{\cal L}(1800)}
\equiv
\frac
{\int{{\cal L} dt (630)}}
{\int{{\cal L} dt (1800)}} 
= 0.934 \pm .060 .\]

\section{$b$ finding Algorithm}

  To identify $b$ hadrons, we begin with a muon as
a seed.  We then select tracks with $p_T > 1.0$~${\rm GeV}/c$ in a cone of
$R \equiv (\Delta \eta^2 + \Delta \phi^2)^{1/2} < 1.0$ and we
require that the invariant mass of the muon-track combination
be below 5.3~${\rm GeV}/c^2$ when the track is assumed to be a pion. 
From this sample we select the track with the highest $p_T$.  The track 
and muon are fit to the constraint that they come from a 
common point.  Events with a fit $\chi^2$-based probability 
greater than 1\% are selected if they also possess a secondary vertex 
within 2~cm of the primary vertex in the transverse plane.  
The number of $b$ hadrons is proportional to the
number of events with the two-track vertex ahead of the primary vertex
(a sample composed of bottom hadrons plus mismeasured tracks)
less the number with the two-track vertex behind the primary vertex
(a sample composed predominantly of mismeasured tracks).  In this
context, ``ahead" means that the secondary vertex displacement $r$ is
in the direction of the momentum vector of the bottom candidate
and ``behind" means that the secondary vertex displacement is
opposite the direction of the momentum vector. We require
the transverse flight distance ($L_{xy} \equiv r \cdot \hat{p_T}$) 
of a $b$ candidate to exceed 250~$\mu{\rm m}$, and the background 
sample to have $L_{xy} < -250~\mu{\rm m}$. Vertices with 
small $|L_{xy}|$ are dominated by prompt particles.

     Monte Carlo studies show that a few percent of real bottom hadrons
are reconstructed in the $-L_{xy}$ sample, that is, behind the primary
vertex.  This exact fraction varies somewhat with different production 
models, most likely because of the different $\Delta R$ and $\Delta \phi$
distributions between the $b$ and $\overline{b}$ hadrons.  However, a
common feature of all Monte Carlos is that this fraction is the
{\em same} at both 630~GeV and 1800~GeV, so the procedure outlined
above still produces an accurate ratio of the number of events produced
from collisions at 630~GeV and 1800~GeV.

     To reduce the contamination in our sample from charm hadrons, 
we require the two-particle mass to be greater than 
1.5~${\rm GeV}/c^2$, where we assume the second track is
a pion.  This is a very tight cut, being at the kinematic 
limit of charm decays, and rejects approximately half 
the $b \rightarrow \mu h^\pm X$ events. There are also indications of a 
high background level (for example, same-sign dimuons) in the low-mass 
sample. ISAJET \cite{ISAJET} Monte Carlo studies show negligible 
charm contamination after these selection requirements.

   This algorithm differs from the ones used in our top quark analyses,
because the algorithms are designed to do quite different 
things: our top quark analysis is designed to identify $b$'s in a 
relatively $b$-poor sample, whereas this algorithm is designed
to accurately count $b$'s in a relatively $b$ rich sample, with
significant $c$-contamination.

\section{Relative Acceptance Calculation}

   To calculate the relative acceptance of the detector at the 
two different energies, two Monte Carlo datasets were created for this 
analysis, one simulating data at $\sqrt{s} = 630$ GeV and the other 
at $\sqrt{s} = 1800$ GeV. Both use the MRSA$^\prime$ parton distributions 
and a renormalization scale of $\mu_0 \equiv \sqrt{m^2_b + p^2_T}$.

   Ten million events were generated at each energy and for a variety
of parton distributions using a $b$-quark Monte Carlo with minimum $b$ quark 
$p_T$ of 6.75~${\rm GeV}/c$ and $|y| \leq 1$ and 
then fragmented using Peterson \cite{Peterson}  fragmentation with
$\epsilon = 0.006$.  The 6.75~${\rm GeV}/c$ point was chosen because
in a sample 10\% this size, no events passing our selection requirements
had a parent $b$ quark with transverse momentum below this value.
Bottom hadron decays were then simulated with version 9.0 of the 
CLEO $B$ Decay Monte Carlo \cite{QQ}, using the standard decay 
tables.  No decays (for instance  $b \rightarrow \mu + X$) were forced, as
$b \rightarrow c \rightarrow \mu$ is about 5\% of the total acceptance
at 630~GeV, and 18\% of the total acceptance at 1800~GeV.  Forcing the
$b$ to decay directly to muons would skew the results.

Events with a muon with a transverse momentum of at least 4.0~${\rm GeV}/c$ 
were then simulated using a fast detector simulation, and events with 
a muon candidate with a transverse momentum of 5.0~${\rm GeV}/c$ or 
greater were kept for further analysis.

    The number of Monte Carlo events in the 1800~GeV sample passing all
cuts is $4045 \pm 67$ after subtraction of events with 
negative $L_{xy}$, and the equivalent number in the 630~GeV 
sample is $2850 \pm 56$.  The
relative acceptance $A_{630}/A_{1800}$ is therefore $0.705 \pm 0.018$.

    A correction to this is necessary as the SVX acceptance in the two
datasets is not identical. The 630~GeV run had the Tevatron's 
final focus running at a nominal $\beta^*$ of approximately 75~cm rather than 
the usual value at 1800~GeV of 35 cm, which widened the $z$ distribution 
of collisions, causing more events to fall outside of the SVX acceptance. 
Additionally, the mean primary vertex position was shifted with 
respect to the 1800~GeV data.

    We measured the acceptance from the data by looking at good CTC
tracks and asking how often a good SVX track is associated with
it.  In particular, we use muons that pass all the cuts in this analysis
although for calculating the acceptance we do not care if they are part
of a $b$ candidate or not.

    We calculate the relative acceptance for the SVX in the following way:

\[ A_{630/1800} = \frac
{(N_\mu(SVX)/N_\mu)_{630}}
{(N_\mu(SVX)/N_\mu)_{1800}} 
\frac{(N^*_\mu(SVX)/N^*_\mu)_{1800}}
{(N^*_\mu(SVX)/N^*_\mu)_{630}} 
.\]

The unstarred quantities are the number of muons in the entire luminous
region, and the starred quantities are the number of muons in the region
where the SVX efficiency and acceptance are at their largest 
(the region where the vertex $z$-position is between 10 and 20~cm on 
both the east and west sides).  This calculation is done to decouple 
the SVX reconstruction probability from the difference in 
acceptance due to the differing beam profile.
This probability may be different for muons from $\pi$ and $K$ decays than
for prompt muons and muons from heavy flavor decays, and the muon
sample composition may differ at the two energies.  This technique
divides out this effect so that only the geometric factor remains.
This approach is equivalent to taking the 1800~GeV SVX 
efficiency curve and superimposing it on the 630~GeV beam profile.  
The measured values are shown in Table~\ref{tab:table1}. We 
calculate a relative acceptance factor due to the beam profile of
$0.817 \pm 0.014$.

\begin{table}
\caption{\label{tab:table1}
Quantities used to determine the 
silicon vertex detector acceptance.}
\begin{ruledtabular}
\begin{tabular}{lcc} 
                             & 1800 GeV & 630 GeV \\ \hline
$N_\mu$ (all)                & 57882  & 28444 \\
$N_\mu$ (with SVX track)     & 37825 & 14213 \\
$N^*_\mu$ (all)              & 13891 & 5913 \\
$N^*_\mu$ (with SVX track)   & 13219 & 5268 \\ 
\end{tabular}
\end{ruledtabular}
\end{table}

    The Monte Carlo dataset used in the acceptance calculation was
generated with $p_T(b) > 6.75$~${\rm GeV}/c^2$ to fully populate the
$p_T$ spectrum, but the convention is to quote the cross section 
above a $p_T(\rm{min})$ such that 90\% of the reconstructed $b$ quarks 
have $p_T > p_T(\rm{min})$.  For this analysis $p_T(\rm{min})$ 
is 10.75~${\rm GeV}/c$.  Because of 
the different $p_T$ spectra at the two energies, Monte Carlo 
datasets that have the same number of entries for 
$p_T(b) > 6.75$~${\rm GeV}/c$ will not have the same 
number of entries for $p_T(b) > 10.75$~${\rm GeV}/c$, so an
additional correction factor of $1.282 \pm 0.007$ is necessary. The uncertainty 
was obtained by varying the scale from $\mu_0$ to $\mu_0/2$ and $2\mu_0$ and 
varying $b$ quark mass from 4.75~${\rm GeV}/c^2$ 
to 4.5 and 5.0~${\rm GeV}/c^2$. Combining all these factors
yields a total relative acceptance $A_{630}/A_{1800} = 0.738 \pm 0.023$.

   A number of studies were made to insure the stability of
this result: we verified that gluon splitting to $c\overline{c}$ 
does not affect this result, nor is it sensitive to the fraction 
of $b$'s produced by gluon splitting rather than $2 \rightarrow 2$ 
processes. We also verified that the algorithm's choice of fragmentation 
tracks over $b$ daughters is the same at both energies. Finally, we verified 
that changing the track selection algorithm leaves the ratio of cross
sections unchanged, and we verified that we were insensitive to the 
value of $b$ quark lifetime when we varied $c \tau$ between 400
and 500 $\mu {\rm m}$.

\section{Results}

\subsection{$b$ Quark Counting}

\begin{figure}
\includegraphics[width=9cm]{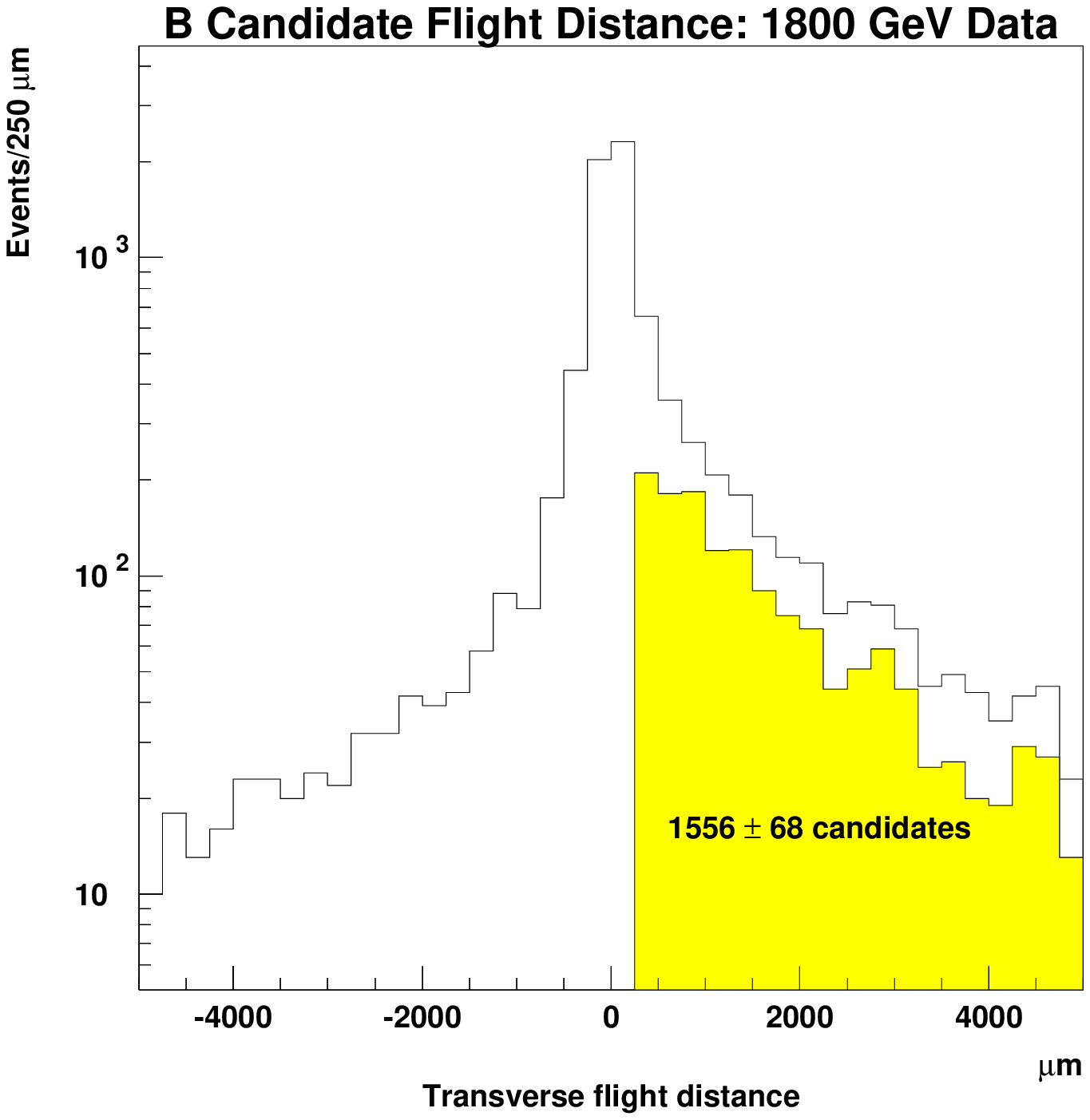}
\caption{The transverse flight distance distribution for $b$ candidates
at $\sqrt{s} = 1800$~GeV.  The shaded region is the excess at large positive
$L_{xy}$.}
\label{Lxy1800}
\end{figure}

\begin{figure}
\includegraphics[width=9cm]{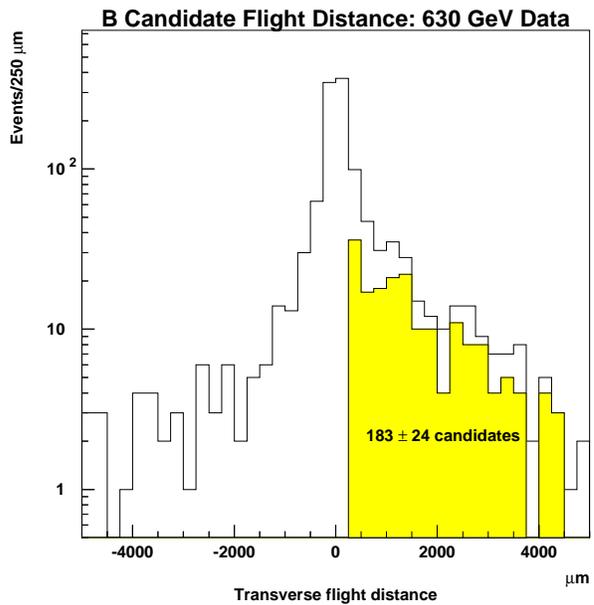}
\caption{The transverse flight distance distribution for $b$ candidates
at $\sqrt{s} = 630$~GeV.  The shaded region is the excess at large positive
$L_{xy}$.}
\label{Lxy0630}
\end{figure}

  The $L_{xy}$ distributions at 1800~GeV and 630~GeV are shown in Figures
\ref{Lxy1800} and \ref{Lxy0630} respectively.  As shown in 
Table~\ref{tab:table2} in the 1800~GeV sample, there are 3083 events at least 
$250~\mu{\rm m}$ ahead of the primary vertex and 1527 events at
least $250~\mu{\rm m}$ behind the primary vertex,
yielding a forward excess of $1556 \pm 68$ events.  In the 630~GeV sample, 
there are 383 events at least $250~\mu{\rm m}$ ahead of the primary 
vertex and 200 events at least $250~\mu{\rm m}$ behind the primary vertex, 
yielding a forward excess of $183 \pm 24$ events.  

\begin{table}
\caption{\label{tab:table2}
Number of candidate events at each energy.}
\begin{ruledtabular}
\begin{tabular}{lcc} 
                                    & 1800 GeV      & 630 GeV \\ \hline
Luminosity ($\rm{nb^{-1}}$)         & $628 \pm 30$  & $582 \pm 24$ \\ \hline
Events $L_{xy} > 250$~$\mu{\rm m}$  & 3083          & 383  \\
Events $L_{xy} < -250$~$\mu{\rm m}$ & 1527          & 200  \\
Forward Excess                      & $1556 \pm 68$ & $183 \pm 24$ \\
\end{tabular}
\end{ruledtabular}
\end{table}

The ratio of observed $b$-quark candidate events (before correcting
for acceptance) is therefore given by:

\[ \frac{N_{630}/{\cal L}_{630}}{N_{1800}/{\cal L}_{1800}}
= 0.126 \pm .020 .\]

\subsection{Relative Cross Section}

   The relative cross section is given by:

\begin{eqnarray}
\frac{\sigma_b(p_T > 10.75)_{630}}{\sigma_b(p_T > 10.75)_{1800}} = &  & \nonumber \\ 
 & \frac{N_b(630)/N_b(1800)}
{A_b(630)/A_b(1800) {\cal L}(630)/{\cal L}(1800)} & \nonumber
\end{eqnarray}
which, when all the factors are put in, yields

\[ \frac{\sigma_b(p_T > 10.75)_{630}}{\sigma_b(p_T > 10.75)_{1800}}
= 0.171 \pm .024 \pm .012 \]

\noindent where the first uncertainty is statistical and the second is
systematic.

\begin{figure}
\includegraphics[width=9cm]{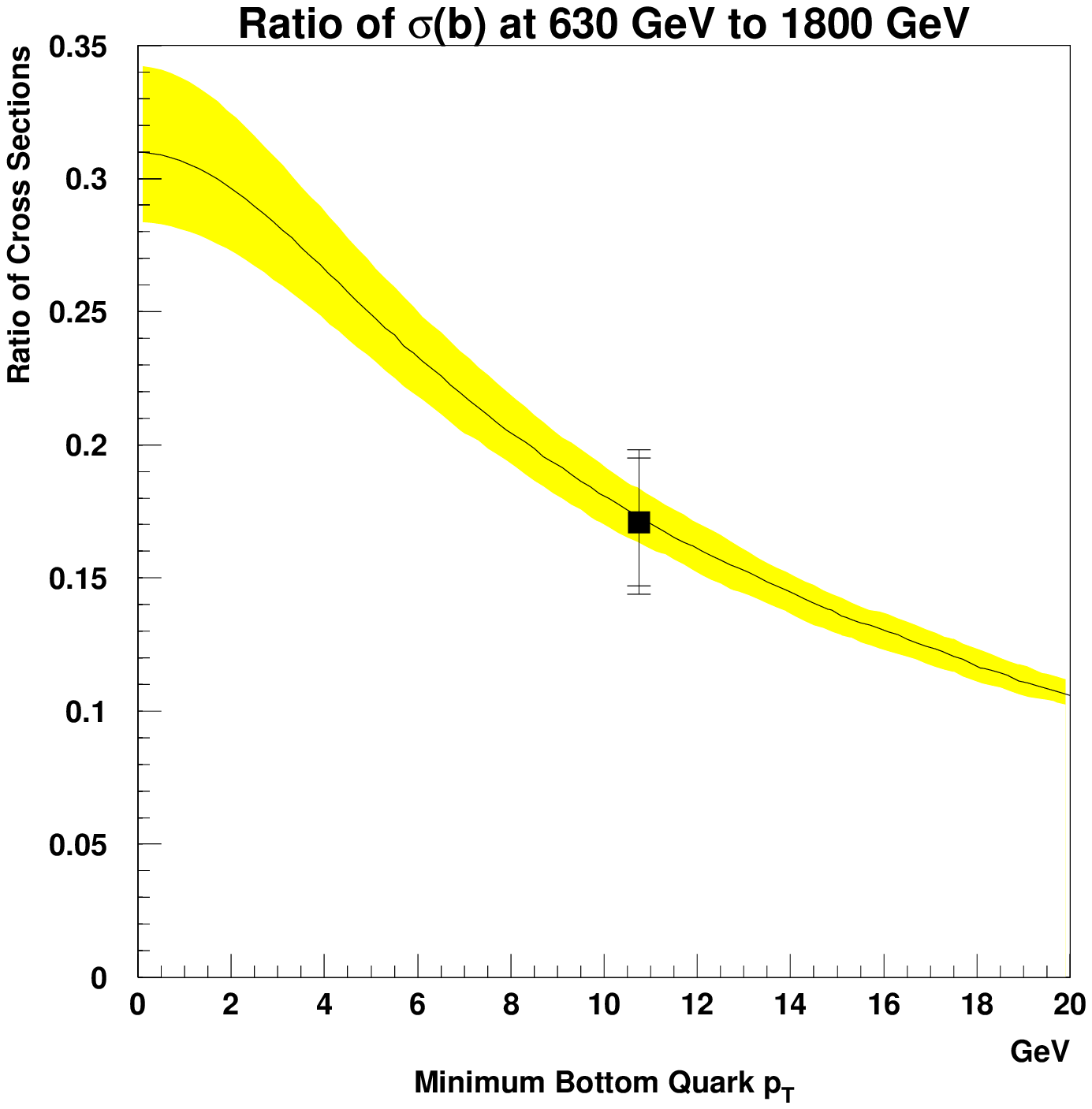}
\caption{The ratio of $\sigma(b)$ 
at $\sqrt{s} = 630$~GeV to $\sqrt{s} = 1800$~GeV 
as a function of the minimum $b$-quark transverse 
momentum, $p_T(\rm{min})$.  The inner error
bars are statistical only, and the outer ones include
systematic uncertainties as well. This is compared to the
NLO QCD prediction using MRSA$^\prime$ parton 
distributions; the central value is obtained with a $b$-quark mass
of 4.75~${\rm GeV}/c^2$ and a renormalization scale of
$\mu_0 = \sqrt{m_b^2+p_T^2}$. The shaded region covers
the variation obtained by varying the scale between $\mu_0/2$ 
and $2\mu_0$ and the mass between 4.5 and 5.0~${\rm GeV}/c^2$. }
\label{ratio_a}
\end{figure}

\begin{figure}
\includegraphics[width=9cm]{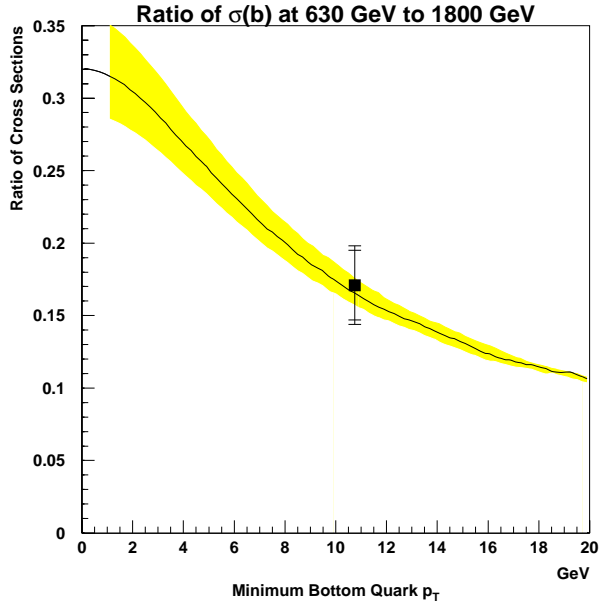}
\caption{The ratio of $\sigma(b)$ 
at $\sqrt{s} = 630$~GeV to $\sqrt{s} = 1800$~GeV 
as a function of the minimum $b$-quark transverse 
momentum, $p_T(\rm{min})$.  The inner error
bars are statistical only, and the outer ones include
systematic uncertainties as well. This is compared to the
NLO QCD prediction using MRST parton 
distributions; the central value is obtained with a $b$-quark mass
of 4.75~${\rm GeV}/c^2$ and a renormalization scale of
$\mu_0 = \sqrt{m_b^2+p_T^2}$. The shaded region covers
the variation obtained by varying the scale between $\mu_0/2$ 
and $2\mu_0$ and the mass between 4.5 and 5.0~${\rm GeV}/c^2$.}
\label{ratio_t}
\end{figure}

The theoretical prediction of NLO QCD \cite{NDE-diff, NDE-tot} 
using MRSA$^\prime$ parton distributions \cite{MRSA} is
$0.174 \pm .011$. The uncertainty was obtained by varying the
renormalization scale from $\mu_0$ to $\mu_0/2$ and $2\mu_0$ and 
by varying the $b$ quark mass from 4.75 ${\rm GeV}/c^2$ 
to 4.5 and 5.0~${\rm GeV}/c^2$.
Our results are compared to NLO QCD predictions using MRSA$^\prime$ 
and MRST \cite{MRST} parton distributions in 
Figures \ref{ratio_a} and \ref{ratio_t} respectively.

\begin{figure}
\includegraphics[width=9cm]{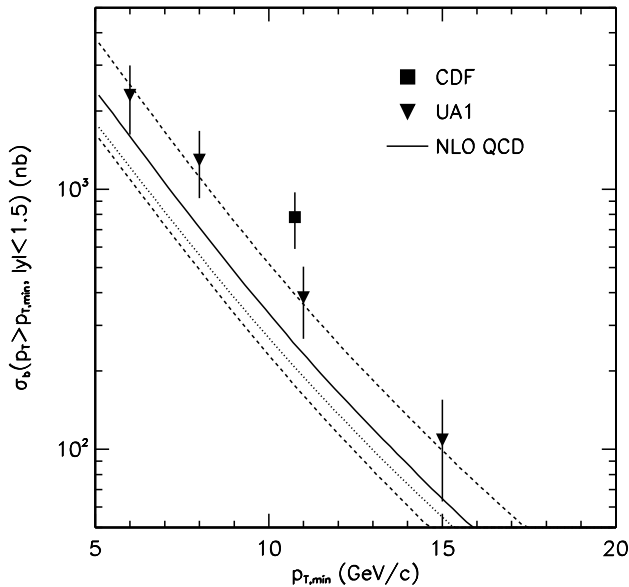}
\caption{The $b$ quark cross section at 630~GeV
for CDF and UA1. The solid line is the 
NLO QCD prediction using MRST parton 
distributions using a renormalization scale of
$\mu_0 = \sqrt{m_b^2+p_T^2}$. The dashed lines cover
a scale variation between $\mu_0/2$ and $2\mu_0$ and a
$b$-quark mass variation between 4.5~${\rm GeV}/c^2$ and 
5~${\rm GeV}/c^2$.
The dotted line is the equivalent of the solid line
except with MRSA$^\prime$ parton distributions.}
\label{b630}
\end{figure}

  We can combine this with our measured $B$ meson cross section
at 1800~GeV \cite{CDFBXsec} and fragmentation ratios \cite{CDFfrag} 
to obtain a cross section at 630~GeV which can be compared 
directly with the results from the UA1 experiment \cite{UA1BXsec}.  
This is shown in Figure~\ref{b630}.  

\section{Conclusions}

     The ratio of the $b$-quark cross sections at 630 and 1800~GeV
matches well with the QCD prediction.  Interpreting this as an
absolute cross-section measurement
at 630~GeV shows our measurement above the UA1 value,
but not so far above that the measurements would be inconsistent at the 95\%
confidence level. The large $b$-quark cross section
is not something that is specific to 1800~GeV data.  It is interesting
to note that NLO QCD predictions using modern parton distributions 
tend to be below the most recent UA1 points as well, although at a level
consistent with their uncertainties.

\begin{acknowledgments}
    We thank the Fermilab staff and the technical staffs of the
participating institutions for their vital contributions.  This work was
supported by the U.S. Department of Energy and National Science Foundation;
the Italian Istituto Nazionale di Fisica Nucleare; the Ministry of Education,
Culture, Sports, Science, and Technology of Japan; the Natural Sciences and 
Engineering Research Council of Canada; the National Science Council of the 
Republic of China; the Swiss National Science Foundation; the A. P. Sloan 
Foundation; the Bundesministerium fuer Bildung und Forschung, Germany; the 
Korea Science and Engineering Foundation (KoSEF); the Korea Research 
Foundation; and the Comision Interministerial de Ciencia y Tecnologia, Spain.
\end{acknowledgments}

\end{document}

%% file: authors.tex

\affiliation{Institute of Physics, Academia Sinica, Taipei, Taiwan 11529, 
Republic of China} 
\affiliation{Argonne National Laboratory, Argonne, Illinois 60439} 
\affiliation{Istituto Nazionale di Fisica Nucleare, 
University of Bologna,I-40127 Bologna, Italy}
\affiliation{Brandeis University, Waltham, Massachusetts 02254} 
\affiliation{University of California at Davis, Davis, California  95616} 
\affiliation{University of California at Los Angeles, Los 
Angeles, California  90024}   
\affiliation{Instituto de Fisica de Cantabria, CSIC-University of Cantabria, 
39005 Santander, Spain} 
\affiliation{Carnegie Mellon University, Pittsburgh, PA  15218} 
\affiliation{Enrico Fermi Institute, University of Chicago, Chicago, 
Illinois 60637} 
\affiliation{Joint Institute for Nuclear Research, RU-141980 Dubna, Russia}

\affiliation{Duke University, Durham, North Carolina  27708} 
\affiliation{Fermi National Accelerator Laboratory, Batavia, Illinois 
60510} 
\affiliation{University of Florida, Gainesville, Florida  32611} 
\affiliation{Laboratori Nazionali di Frascati, Istituto Nazionale di Fisica
Nucleare, I-00044 Frascati, Italy} 
\affiliation{University of Geneva, CH-1211 Geneva 4, Switzerland} 
\affiliation{Glasgow University, Glasgow G12 8QQ, United Kingdom}
\affiliation{Harvard University, Cambridge, Massachusetts 02138} 
\affiliation{Hiroshima University, Higashi-Hiroshima 724, Japan} 
\affiliation{University of Illinois, Urbana, Illinois 61801} 
\affiliation{The Johns Hopkins University, Baltimore, Maryland 21218} 
\affiliation{Institut f\"{u}r Experimentelle Kernphysik, 
Universit\"{a}t Karlsruhe, 76128 Karlsruhe, Germany} 
\affiliation{Center for High Energy Physics: Kyungpook National
University, Taegu 702-701; Seoul National University, Seoul 151-742; and
SungKyunKwan University, Suwon 440-746; Korea} 
\affiliation{High Energy Accelerator Research Organization (KEK), Tsukuba, 
Ibaraki 305, Japan} 
\affiliation{Ernest Orlando Lawrence Berkeley National Laboratory, 
Berkeley, California 94720} 
\affiliation{Massachusetts Institute of Technology, Cambridge,
Massachusetts  02139}    
\affiliation{University of Michigan, Ann Arbor, Michigan 48109} 
\affiliation{Michigan State University, East Lansing, Michigan  48824} 
\affiliation{University of New Mexico, Albuquerque, New Mexico 87131} 
\affiliation{Northwestern University, Evanston, Illinois  60208} 
\affiliation{The Ohio State University, Columbus, Ohio  43210} 
\affiliation{Osaka City University, Osaka 588, Japan} 
\affiliation{University of Oxford, Oxford OX1 3RH, United Kingdom} 
\affiliation{Universita di Padova, Istituto Nazionale di Fisica 
Nucleare, Sezione di Padova, I-35131 Padova, Italy} 
\affiliation{University of Pennsylvania, Philadelphia, Pennsylvania 19104}    
\affiliation{Istituto Nazionale di Fisica Nucleare, University and Scuola
Normale Superiore of Pisa, I-56100 Pisa, Italy} 
\affiliation{University of Pittsburgh, Pittsburgh, Pennsylvania 15260} 
\affiliation{Purdue University, West Lafayette, Indiana 47907} 
\affiliation{University of Rochester, Rochester, New York 14627} 
\affiliation{Rockefeller University, New York, New York 10021} 
\affiliation{Rutgers University, Piscataway, New Jersey 08855} 
\affiliation{Texas A\&M University, College Station, Texas 77843} 
\affiliation{Texas Tech University, Lubbock, Texas 79409} 
\affiliation{Institute of Particle Physics, University of Toronto, Toronto
M5S 1A7, Canada} 
\affiliation{Istituto Nazionale di Fisica Nucleare, University of Trieste/
Udine, Italy} 
\affiliation{University of Tsukuba, Tsukuba, Ibaraki 305, Japan} 
\affiliation{Tufts University, Medford, Massachusetts 02155} 
\affiliation{Waseda University, Tokyo 169, Japan} 
\affiliation{University of Wisconsin, Madison, Wisconsin 53706} 
\affiliation{Yale University, New Haven, Connecticut 06520}


\author{D.~Acosta}
\affiliation{University of Florida, Gainesville, Florida  32611} 

\author{T.~Affolder}
\affiliation{Ernest Orlando Lawrence Berkeley National Laboratory, 
Berkeley, California 94720} 

\author{H.~Akimoto}
\affiliation{Waseda University, Tokyo 169, Japan} 

\author{M.~G.~Albrow}
\affiliation{Fermi National Accelerator Laboratory, 
Batavia, Illinois 60510} 

\author{D.~Ambrose}
\affiliation{University of Pennsylvania, Philadelphia, 
Pennsylvania 19104}  

\author{D.~Amidei}
\affiliation{University of Michigan, Ann Arbor, Michigan 48109} 

\author{K.~Anikeev}
\affiliation{Massachusetts Institute of Technology, 
Cambridge, Massachusetts  02139} 

\author{J.~Antos}
\affiliation{Institute of Physics, Academia Sinica, 
Taipei, Taiwan 11529, Republic of China} 

\author{G.~Apollinari}
\affiliation{Fermi National Accelerator Laboratory, 
Batavia, Illinois 60510} 

\author{T.~Arisawa}
\affiliation{Waseda University, Tokyo 169, Japan} 

% ...

\collaboration{The CDF Collaboration}

%% file: b630.bbl
\begin{thebibliography}{99}
\bibitem{Phi} In the CDF coordinate system, $\theta$ and $\phi$ are the polar
and azimuthal angles, respectively, defined with respect to the proton beam
direction $z$.  The pseudorapidity $\eta$ is defined as $- \ln \tan(\theta/2)$.  
The transverse momentum of a particle is $p_T = p \sin (\theta)$.
\bibitem{luminosity}  D.~Cronin-Hennessy, A.~Beretvas and P.~F.~Derwent,
Nucl.\ Instrum.\ Meth.\ A {\bf 443}, 37 (2000).  Note that
this uncertainty is somewhat smaller than what is usually 
quoted by CDF.  This is because this is only the component of the 
luminosity uncertainty that is not in common between the 
measurements at 1800~GeV and 630~GeV.
\end{thebibliography}
